\documentclass[%
 reprint,
 superscriptaddress,
 amsmath,amssymb,
 aps,
 prd,
]{revtex4-2}

\usepackage{graphicx}
\usepackage{dcolumn}
\usepackage{bm}
\usepackage{hyperref}
\usepackage{xcolor}

\begin{document}

\newcommand{\rv}[1]{\mathbf{#1}} 
\newcommand{\uv}[1]{\hat{\mathbf{#1}}} 
\newcommand{\fv}[1]{\mathsf{#1}} 
\newcommand{\mset}[1]{\mathbb{#1}} 
\newcommand{\mat}[1]{\mathbf{#1}} 
\newcommand{\tens}[1]{\mathbb{#1}} 
\newcommand{\spinor}[1]{\mathsf{#1}} 
\newcommand{\vp}{\wedge} 

\newcommand{\pderiv}[2]{\frac{\partial #1}{\partial #2}}
\newcommand{\pderivn}[3]{\frac{\partial^{#1}#2}{\partial {#3}^{#1}}}
\newcommand{\fderiv}[2]{\frac{d #1}{d #2}}
\newcommand{\fderivn}[3]{\frac{d^{#3} #1}{d {#2}^{#1}}}

\newcommand{\braket}[1]{\langle #1 \rangle}

\preprint{APS/123-QED}

\title{Data-driven core collapse supernova multilateration with first neutrino events}

\author{Farrukh Azfar} 
\author{Jeff Tseng} 
\affiliation{%
 Department of Physics, Oxford University, Oxford OX1 3RH United Kingdom
}%
\author{Marta Colomer Molla} 
\affiliation{Universit\'e Libre de Bruxelles, 1050 Bruxelles, Belgium}
\author{Kate Scholberg} 
\affiliation{Department of Physics, Duke University, Durham, North Carolina 27708, USA}
\author{Alec Habig}
\affiliation{Department of Physics and Astronomy, University of Minnesota Duluth, Duluth,
Minnesota 55812, USA}
\author{Segev BenZvi}
\affiliation{Department of Physics and Astronomy, University of Rochester, Rochester,
New York 14627, USA}
\author{Melih Kara}
\affiliation{Institute for Astroparticle Physics, Karlsruhe Institute of Technology,
76021 Karlsruhe, Germany}
\author{James Kneller}
\affiliation{Department of Physics, North Carolina State University, Raleigh,
North Carolina 27695, USA}
\author{Jost Migenda}
\affiliation{e-Research, King's College London, London, United Kingdom}
\author{Dan Milisavljevic}
\affiliation{Department of Physics and Astronomy, Purdue University, West Lafayette,
Indiana 47907, USA}
\author{Evan O'Connor}
\affiliation{The Oskar Klein Centre, Department of Astronomy, Stockholm University,
AlbaNova, SE-10691 Stockholm, Sweden}

\date{\today}

\begin{abstract}
A Galactic core-collapse supernova (CCSN) is likely to be observed in
neutrino detectors around the world minutes to hours before the electromagnetic radiation arrives. The SNEWS2.0 network of neutrino and dark matter detectors aims to use the relative arrival times of the neutrinos at the different experiments to point back to the supernova
so as to facilitate follow-up observation. One of the simplest methods to estimate the CCSN direction is to use the first neutrino events detected through the inverse beta decay (IBD) process, $\overline{\nu}_e p\rightarrow e^+n$. We will consider neutrino detectors sensitive to IBD interactions with low backgrounds. The difference in signal arrival times between a large and a small detector will be biased, however,
with the first event at the smaller detector, on average, arriving later than that at the larger detector.
This bias can be mitigated by using these first events in a data-driven approach without recourse to simulations or models.
The resulting method requires, at minimum,
only the times of the first events at most detectors, along
with a longer time series of events from one larger detector to act as a reference lightcurve.
In this article, we demonstrate this method and its uncertainty estimate using pairs of detectors of different sizes and with different supernova distances.
Finally, we use this method to calculate probability skymaps using four detectors
currently in operation (Super-Kamiokande, JUNO, LVD, and SNO+) and show that the calculated probabilities yield appropriate confidence intervals for all supernova directions.  The area of the 68\% confidence interval varies by
distance and direction, but is expected to be a few thousand square degrees.
The resulting skymaps should be useful for the multi-messenger community
as a rapid, initial pointing to follow up on the SNEWS2.0 Galactic CCSN neutrino alert.

\end{abstract}

\maketitle


\section{\label{sec:intro} Introduction}

The collapse of a massive star in the Milky Way is
a unique opportunity rich with physics
on topics spanning astrophysics, cosmology, nuclear physics, and fundamental particle physics\footnote{See, for instance, \cite{koshio2022snowmass2021topicalgroup,Furusawa:2022ktu,Arcones:2016euo}, and recent articles
such as \cite{visinelli2024neutrinosbendconsequencesultralight,lella2024supernovalimitsqcdaxionlike,li2024lowenergysupernovaconstraintsmillicharged}.}. At the same time, it is an exceedingly rare phenomenon, with rates estimated to be
around $1.63\pm 0.46$ per century~\cite{Rozwadowska:2020nab}. In order to take maximal
advantage of the next Galactic core-collapse supernova (CCSN), the upgrade of the
SuperNova Early Warning System
(SNEWS2.0)~\cite{SNEWS:2020tbu}
aims to utilize the burst of neutrinos, which precedes electromagnetic radiation by
minutes to hours, to alert the world's observers. The SNEWS2.0 network also aims to use the relative arrival times of the neutrino bursts at each detector to point back to the supernova and guide follow-up observations.

The idea of multilateration (or, informally, ``triangulation'') using the neutrino arrival
times has been explored by multiple
authors~\cite{Beacom:1998fj,Muhlbeier:2013gwa,Brdar:2018zds,Linzer:2019swe,Coleiro:2020vyj}.
One of the simplest methods is to use the first observed events to determine
the time differences between detector observations.  For this method to be effective, the background should be negligible, which is a reasonable assumption for many underground detectors on the approximately ten-second timescale of a supernova burst.  In addition to its simplicity, the method has the added advantages of being
fast as well as relatively independent of supernova models,
and thus works in almost all circumstances.
It is expected, of course, that pointing accuracy will ultimately be dominated by neutrino-electron
elastic scattering events in large detectors such as Super-Kamiokande~\cite{Super-Kamiokande:2024pmv}
and DUNE~\cite{DUNE:2024ptd}, which expect pointing precisions between 3 and 7 degrees
for a distance of 10~kpc, depending on supernova model,
but simple multilateration through a combination of detectors can provide added quickness and robustness to the search for multi-messenger counterparts.

In this article, we describe a refined first-event method which reduces the inherent
bias due to comparing detectors with different event yields, along with a practical data-driven
method for estimating directional uncertainty, which is critical for assigning confidence levels to
different directions (Sec.~\ref{sec:method}).  Sec.~\ref{sec:sim} describes the very
simple simulation setup used to evaluate the method's performance, and Sec.~\ref{sec:results}
presents the
results of that evaluation.  We combine multiple detectors in Sec.~\ref{sec:pointing} to
make skymaps with confidence intervals, and discuss the results in Sec.~\ref{sec:discussion}.

\section{\label{sec:method} Method}

Multilateration involves calculating the most likely direction to the
supernova using the differences between arrival times of the neutrino bursts
at different detectors.  In principle, if the direction to the supernova is
given by the unit vector $\uv{n}$, we can predict the arrival time
differences (the ``true lags'') between two detector positions $\rv{r}_A$ and $\rv{r}_B$ using
the formula
\begin{equation}
\tau_{AB}(\uv{n}) = - \frac{(\rv{r}_A-\rv{r}_B)\cdot\uv{n}}{c},
\label{eq:lag}
\end{equation}
where we have assumed that the neutrinos travel at the speed of light $c$.
If we measure corresponding time differences $\Delta t_{AB}$, we can find
the most likely direction by varying $\uv{n}$ and minimizing the quantity
\begin{equation}
\chi^2(\uv{n}) = \sum_{AB}\sum_{CD} (\Delta t_{AB} - \tau_{AB}(\uv{n})) V^{-1}_{AB,CD}
(\Delta t_{CD} - \tau_{CD}({\uv{n})})
\label{eq:chi2}
\end{equation}
where the sums are over detector pairs $AB$ and $CD$, and $V_{AB,CD}$
is the covariance matrix of the time differences.
(We will use $\tau_{AB}(\uv{n})$ to denote the true lag given an arbitrary direction $\uv{n}$,
while $\tau_{AB}$ with no explicit dependence will refer to the true lag given
the true supernova direction.)
It should be obvious that it is important to reduce any biases in determining
the time differences $\Delta t_{AB}$, as they may guarantee that follow-up observers
will be watching the wrong patch of sky for the explosion.

Equally important for multilateration is the estimation of the covariance matrix $V$,
as it determines the size and shape of the confidence intervals, and therefore the
size and priority of search areas.
Previous methods~\cite{Brdar:2018zds,Linzer:2019swe,Coleiro:2020vyj} have
generally relied on evaluating such uncertainties by running Monte Carlo
trials with either model- or data-derived inputs.  Such methods are
potentially time consuming, and in the case of toy models, may introduce
further systematic uncertainties.  The other main aim of this paper is
to estimate these uncertainties using direct calculations from the data
itself.

For this paper, we will focus on four geographically dispersed detectors
which are currently in operation and
are primarily sensitive to the inverse beta decay (IBD)
process, $\overline{\nu}_e p\rightarrow e^+n$:
Super-Kamiokande (SK, Japan)~\cite{Super-Kamiokande:2002weg},
JUNO (China)~\cite{JUNO:2021vlw},
LVD (Italy)~\cite{Aglietta:1992dy},
and SNO+ (Canada)~\cite{SNO:2021xpa}.
The IBD channel possesses advantages of sensitivity
to a single anti-neutrino flavor, a large interaction cross section, and accessibility in
a number of operating water Cherenkov and liquid scintillator detectors.
Moreover, in detectors which can tag the subsequent neutron capture on protons or,
in the case of Super-Kamiokande, on gadolinium nuclei, non-IBD events can be reduced significantly;
for this study we assume the background is negligible.
We assume further that the neutrino event rates as a function of time observed in every detector differ in their overall normalization but are otherwise nearly identical in shape, {\it i.e.}, the neutrino events are well above
trigger thresholds, and Earth-matter effects are negligible.
This assumption is not strictly true for this set of detectors, since
Super-Kamiokande applies a 7~MeV energy threshold on
the prompt positron; while this threshold is well below the mean energy of 20~MeV
expected for observed supernova $\overline{\nu}_e$'s, it introduces a small and mostly
negligible delay to its time distribution,
which is not taken into account in this method.
Finally, we neglect the very small, relatively low energy flux of $\overline{\nu}_e$,
indicative of late-stage nuclear burning,
which is expected to precede core collapse~\cite{Odrzywolek:2003vn,Odrzywolek:2004em,Patton:2015sqt,Patton:2017neq}.

\subsection{\label{sec:bias} First-event bias}

As noted above, one of the simplest methods to calculate the
$\Delta t_{AB}$ is to 
take the time difference between the first events observed at each detector.
This method, however, incurs a significant bias due to the fact that a larger
detector will in general see its first event earlier, relative to the beginning
of the burst, than a smaller detector.  In this section, we derive a method
for mitigating this bias.

First, we derive a numerical estimate of the expectation value of
the first event time in one detector.
Let $t_j$ be the time of the $j$-th observed event at a single detector,
with $j=1, 2, \cdots N$.
The probability density function (PDF)
for observing the first event at time $t_1$ is the product of two probabilities: first, of observing zero events in the time interval $(-\infty,t_1)$ followed by a single event in $[t_1,t_1+dt_1)$. The Poisson PDF of the first is simply $e^{-\mu(t_1)}$, where $\mu(t_1)=\int_{-\infty}^{t_1}R(t)dt$
and $R(t)$ is the event rate at time $t$, {\em i.e.},
the number of events in the interval $[t,t+dt)$.
The PDF of the second is simply $R(t_1)$, leading to the first-event PDF~\cite{Brdar:2022vfr}
\[ p_1(t_1) = R(t_1)e^{-\mu(t_1)}. \]
The expectation value of the first event time is then
\begin{equation}
E(t_1)
= \frac{\int_{-\infty}^\infty t p_1(t) dt}{\int_{-\infty}^\infty p_1(t) dt}
= \frac{\int_{-\infty}^\infty tR(t)e^{-\mu(t)} dt}{
        \int_{-\infty}^\infty R(t)e^{-\mu(t)} dt}.
\label{eq:expt1}
\end{equation}

We estimate $E(t_1)$ numerically by
using the observed event times $t_j$ in the following manner:
we treat the $t_j$ as random variables which are independent (up to the fact
that they are sorted), and we approximate the integrals in Eq.~\ref{eq:expt1} in
the style of a Monte Carlo integration:
\begin{eqnarray*}
\int_{-\infty}^\infty tR(t)e^{-\mu(t)} dt & \approx &
   \sum_{j=1}^N t_j e^{-\mu(t_j)} \approx \sum_{j=1}^N t_j e^{-j} \\
\int_{-\infty}^\infty R(t)e^{-\mu(t)} dt & \approx &
   \sum_{j=1}^N e^{-\mu(t_j)} \approx \sum_{j=1}^N e^{-j}.
\end{eqnarray*}
In effect, in the first approximation,
we treat $R(t)$ in the integrand as a weight function,
with larger values indicating a higher density of events.
In the second relation, we
approximate the cumulative distribution $\mu(t_j)$ with the
number of events up to that time, which is simply $j$.
The numerical estimate of $E(t_1)$, which we denote $\braket{t_1}$, is thus
\begin{equation}
E(t_1) \approx \braket{t_1} \equiv \frac{\sum_{j=1}^N t_j e^{-j}}{\sum_{j=1}^N e^{-j}},
\label{eq:nexpt1}
\end{equation}
which is essentially a weighted average which emphasizes the early events
and exponentially suppresses the later ones.
Indeed, it is clear that one needs only the early events to calculate these
sums to an appropriate precision.
On the other hand, the approximations used above can introduce their own biases;
for instance, it is easily shown that $\braket{t_1} > t_1$ by construction,
whereas the first event time
could occur before or after the true expectation value $E(t_1)$.
It can also be seen that the approximation $\mu(t_j)\approx j$ works best for early times if the
rate rises quickly.  Conversely, if the rate rises slowly, the approximation can
introduce sizable relative deviations from $\mu(t_j)$.
It is therefore worth noting that
there is good agreement between supernova modeling groups on the rapid rise of the
$\overline{\nu}_e$ lightcurve at early times---the channel and phase most relevant to this
study~\cite{OConnor:2018sti,Cabezon:2018lpr}.

If we wish to use the events of a large detector to estimate $\braket{t_1}$ for
a smaller detector, we can rescale the larger detector's rate $R(t)$
by a constant factor $\alpha<1$, such that
\[ R'(t) = \alpha R(t). \] 
The rescaled estimate corresponding to Eq.~\ref{eq:nexpt1} then becomes
\begin{equation}
E(t_1^\alpha) \approx
\braket{t_1^\alpha} \equiv \frac{\sum_{j=1}^N t_j e^{-\alpha j}}{\sum_{j=1}^N e^{-\alpha j}}.
\label{eq:nexpt1a}
\end{equation}
The difference between $\braket{t_1}$ and $\braket{t_1^\alpha}$ is the estimated bias of the observed first-event arrival times between two detectors of unequal size.

Compared to Eq.~\ref{eq:nexpt1}, the exponential suppression in Eq.~\ref{eq:nexpt1a}
is smaller term by term, and thus can be seen as a weighted average farther into the time series.
In fact, it can be proven (see App.~\ref{app:alpha}) that
$\braket{t_1^\alpha} > \braket{t_1}$
for $\alpha<1$.
It is also evident that when calculating $\braket{t_1^\alpha}$, one should increase
the number of early events by a factor of $\alpha^{-1}$ in order to achieve
a similar desired level of accuracy.

Now consider two detectors $A$ and $B$ which receive the neutrino signal with a true lag
$\tau_{AB}$
between them.  Experiment $A$ observes $N_A$ events, $\{t^A_1, ..., t^A_{N_A}\}$, while experiment $B$ receives
$N_B=\alpha N_A$ events, $\{t^B_1, ..., t^B_{N_B}\}$.  We define the ``corrected lag'' as
\begin{equation}
    Z_{AB} \equiv t^A_1-t^B_1 - \braket{t_1^A} + \braket{t_1^{\alpha A}},
\label{eq:z}
\end{equation}
where $\braket{t_1^{\alpha A}}$ indicates the rescaled estimate from Eq.~\ref{eq:nexpt1a},
based on the event times from detector $A$.  Crucially, $\braket{t_1^A}$ and
$\braket{t_1^{\alpha A}}$ have no lag between them; the only difference in the
expectation values is due to the relative yield $\alpha$.
The difference $\braket{t_1^A}-\braket{t_1^{\alpha A}}$
therefore represents the estimated bias in the first-event difference
due only to the relative yields
of the two detectors.  The expectation value of $Z_{AB}$ is thus, ideally,
the true lag itself,
{\it i.e.}, $\tau_{AB}$ evaluated with the true supernova direction.
We call $Z_{AB}-\tau_{AB}$ the ``residual bias'',
and we use $Z_{AB}$ as our estimate of $\Delta t_{AB}$ in Eq.~\ref{eq:chi2}.

\subsection{Estimated uncertainty}

The expected uncertainty on $Z_{AB}$ is determined by noting that $t_1^A$ and $t_1^B$ are independent
measurements and hence
\begin{equation}
    \sigma_{Z_{AB}}^2 = \sigma_{A}^2 + \sigma_{B}^2,
    \label{eq:varz}
\end{equation}
where $\sigma_A$ and $\sigma_B$ are the uncertainties on $t_1^A$ and $t_1^B$,
respectively.
The variance in $t_1^A$ requires estimating the expectation value,
following Eq.~\ref{eq:expt1},
\[
E((t_1^A)^2)
= \frac{\int_{-\infty}^\infty t^2 p_1(t) dt}{\int_{-\infty}^\infty p_1(t) dt}
\approx \braket{(t_1^A)^2}
\equiv \frac{\sum_{j=1}^{N_A} (t_j^A)^2 e^{-j}}{\sum_{j=1}^{N_A} e^{-j}}.
\]
The variance in $t_1^B$, which is invariant with respect to the lag,
can be evaluated using the data of detector $B$, or rescaling
the data of detector $A$:
\begin{eqnarray*}
E((t_1^B-\braket{t_1^B})^2) & \approx & \braket{(t_1^B)^2} - \braket{t_1^B}^2 \\
& \approx & \braket{(t_1^{\alpha A})^2} - \braket{t_1^{\alpha A}}^2.
\end{eqnarray*}
In practice, since the events in both detectors are subject to statistical
fluctuations, we evaluate the expectation value with both datasets, and take the larger variance
as a conservative estimate of $\sigma^2_{B}$.

If we calculate all time differences with respect to a single ``reference'' detector,
which we designate as $A$, the covariance matrix of Eq.~\ref{eq:chi2} takes the simple form
\begin{equation*}
V =
    \left(\begin{array}{cccc}
    \sigma_A^2+\sigma_B^2 & \sigma_A^2 & \sigma_A^2 & \cdots \\
    \sigma_A^2 & \sigma_A^2+\sigma_B^2 & \sigma_A^2 & \\
    \sigma_A^2 & \sigma_A^2 & \sigma_A^2+\sigma_C^2 & \\
    \vdots & & & \ddots
    \end{array}\right).
\end{equation*}
Eq.~\ref{eq:chi2} then becomes
\begin{equation}
    \chi^2(\uv{n}) = \sum_B \sum_C (\Delta t_{AB}-\tau_{AB}(\uv{n}))V^{-1}
    (\Delta t_{AC}-\tau_{AC}(\uv{n}))
    \label{eq:chi2single}
\end{equation}
where the sums are over all the detectors besides $A$.  This formula is used in
Sec.~\ref{sec:pointing} to evaluate directional confidence intervals with simulated supernovae.

\subsection{Binned data}

It is possible, though it will not be implemented for the studies in this article,
to derive a similar technique to estimate $E(t_1^\alpha)$ from data
which is binned in time, and may include significant background levels.
For instance, consider data from experiment $A$ reported as $M$ event counts $n_k$
in time windows $[T_k,T_{k+1})$, with an estimated background $b_k$ in
each bin.  The time windows do not have to be uniform in length, but they should be
long enough such that $n_k-b_k>0$.
Eq.~\ref{eq:nexpt1} then becomes
\begin{equation}
    E(t_1) \approx
    \braket{t_1^\alpha} = \frac{\sum_{k=1}^M (n_k-b_k) T_k e^{-\alpha\mu_k}}{
    \sum_{k=1}^M (n_k-b_k) e^{-\alpha\mu_k}}.
    \label{eq:nexpt1bin}
\end{equation}
where
\begin{equation*}
    \mu_k=\sum_{r=1}^{k-1}(n_r-b_r)
\end{equation*}
is the background-subtracted sum of event counts before time $T_k$.
Similar expressions can be obtained to provide an estimate of the variance.
In this way, neutrino telescopes such as IceCube~\cite{IceCube:2016zyt,IceCube:2021mpg}
and KM3NeT~\cite{KM3Net:2016zxf}, which have large
IBD signal yields but also large backgrounds and cannot identify CCSN neutrino events individually, can be incorporated directly
into this supernova multilateration technique. In fact, it is expected that IceCube and KM3NeT will fit their data to determine the start of their neutrino bursts, and will report those times along with their own estimates of the uncertainty, as reported in ~\cite{Cross:2019jpb,KM3NeT:2021moe}.

\section{\label{sec:sim} Simulation}

For the purpose of generating event sequences with which to test our method,
we utilize the IBD lightcurves of the widely-employed
{\tt Bollig\_2016}~\cite{Mirizzi:2015eza}
models of SNEWPY~\cite{Baxter_2022,Baxter2021},
assuming a progenitor with $27M_\odot$ and the LS220 equation of state~\cite{Woosley:2007as}.
We also present results using a $11.2M_\odot$ progenitor mass,
with the same equation of state, for test purposes.  

Unless mentioned otherwise, for
all the Monte Carlo trials in this paper, we use the same, but otherwise arbitrary,
benchmark supernova time (1 Nov 2021 at 05:22:36.328 GMT)
and direction ($300^\circ$ right ascension and $-30^\circ$ declination).
The four detectors are modelled using
SNOwGLoBES\footnote{A. Beck {\it et al.}, \texttt{https://github.com/SNOwGLoBES/snowglobes}, 2016}
assuming a distance of 10~kpc, adiabatic MSW oscillations, and normal neutrino
mass ordering.
The true lag times among the detectors
are given in Table~\ref{tab:truelags}.
\begin{table}
    \begin{center}
        \begin{tabular}{|l|l|r|}
        \hline
        Detector 1 & Detector 2 & True lag (ms) \\
        \hline
        Super-Kamiokande & JUNO & -1.97 \\ 
        Super-Kamiokande & LVD & -25.15 \\ 
        Super-Kamiokande & SNO+ & -14.66 \\ 
        JUNO & LVD & -23.17 \\ 
        JUNO & SNO+ & -12.69 \\ 
        LVD & SNO+ & 10.48 \\ 
        \hline
        \end{tabular}
    \end{center}
    \caption{True lag times ($\tau$) for the neutrino bursts between pairs of detectors,
    for the simulated supernova.}
    \label{tab:truelags}
\end{table}

\section{\label{sec:results} Results}

\subsection{\label{sec:before} Before bias correction}

\begin{figure}
    \includegraphics[width=3.4in]{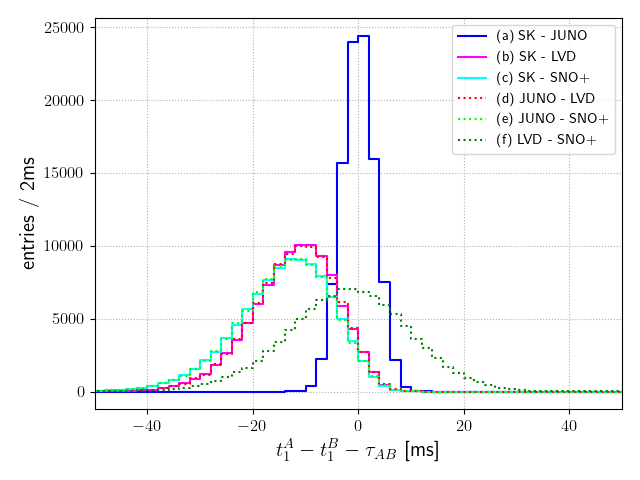}
    \caption{Before bias correction:  distributions of first-event time differences (observed lags) less true lags,
    $t_1^A-t_1^B-\tau_{AB}$,
    over 100,000 trials with the $27M_\odot$ IBD lightcurve shape at 10~kpc.
    Pairwise differences are always computed such that detector $A$ is larger.}
    \label{fig:prelags}
\end{figure}

We run 100,000 Monte Carlo trials,
each including generated IBD events with average yields given in Table~\ref{tab:yields}, following \cite{SNEWS:2020tbu}.
Fig.~\ref{fig:prelags} shows the comparison between the observed first-event time differences, $t_1^A-t_1^B$, before any bias correction, and the true lag $\tau_{AB}$,
for all pair combinations of the four detectors.
As expected, experiments with similar yields (Super-Kamiokande with JUNO and LVD with SNO+) are centered close to zero, indicating small bias,
while pairings of large detectors with small detectors exhibit significant bias.
The distributions tend to peak on the negative side, because we have taken detector $A$ always to be the larger one, which on average observes the first event earlier.
\begin{table}
    \begin{tabular}{|l|r|r|}
    \hline
    Detector & $27M_\odot$ yield & $11.2M_\odot$ yield \\
    \hline
    Super-Kamiokande & 7800 & 4000 \\
    JUNO & 7200 & 3800 \\
    LVD & 360 & 190 \\
    SNO+ & 280 & 150 \\
    \hline
    \end{tabular}
    \caption{Average IBD event yields assuming a distance of 10~kpc, adiabatic MSW oscillations, and normal neutrino mass ordering,
    from \cite{SNEWS:2020tbu}.}
    \label{tab:yields}
\end{table}

\begin{figure}
    \includegraphics[width=3.4in]{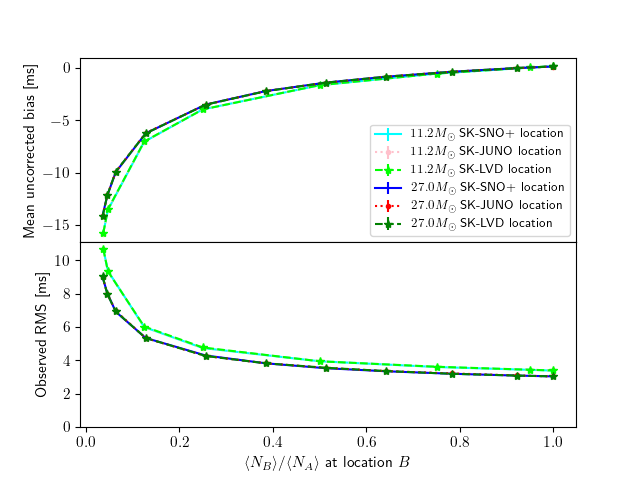}
    \caption{Top:
    mean uncorrected bias as a function of relative detector yield
    for Super-Kamiokande compared with detector locations
    at JUNO (red/pink), LVD (green/lime green), and SNO+ (blue/cyan).
    For each value of $\braket{N_B}/\braket{N_A}$, 100,000 Monte Carlo trials have
    been run with IBD lightcurves for two progenitor masses
    ($27M_\odot$ and $11.2M_\odot$).  In each case, the average Super-Kamiokande
    yield is held constant at the value in Table~\ref{tab:yields}.
    Bottom:  mean observed RMS of the uncorrected distributions.
    For both figures, the differences between curves for the same
    progenitor are less than 0.1~ms.}
    \label{fig:prebiasyield}
\end{figure}

We further examine the effect of relative detector yields before any correction
by running Monte Carlo
trials in which the average Super-Kamiokande yield is held constant,
while the average yield is
varied at the other detector locations.  
Fig.~\ref{fig:prebiasyield} shows the resulting mean bias
as a function of the relative
average yield between the smaller and larger detectors, for the two progenitor masses.
For the smallest relative yields, comparable to that of pairing Super-Kamiokande
with SNO+, the uncorrected bias can exceed 10~ms, while the bias
approaches zero as the detector yields become more equal (note that for equal yields,
the bias is actually slightly positive, around $+0.2$~ms, because of the higher energy
threshold at Super-Kamiokande).  There are small differences between the two progenitors,
but for the same progenitor, the bias and its observed RMS are
largely independent of the detector locations.

\begin{figure}
    \includegraphics[width=3.4in]{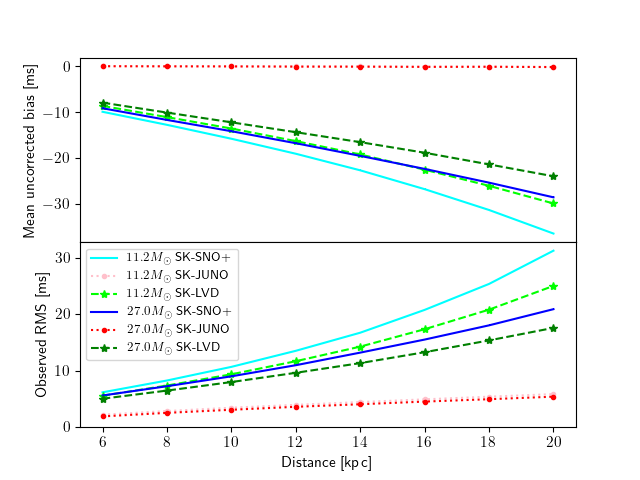}
    \caption{Top:
    mean uncorrected bias for Super-Kamiokande compared with other detectors
    at different distances, using Table~\ref{tab:yields} for the yields at 10~kpc.
    Bottom:  mean observed RMS of the uncorrected distributions.}
    \label{fig:prebiasdistance}
\end{figure}

In order to examine the effect of varying the overall event yields,
we run Monte Carlo trials with nominal yields scaled as if the
model supernova has been moved to different distances, from 6 to 20~kpc.
The mean uncorrected biases and their mean observed RMS are shown as a function of
supernova distance in Fig.~\ref{fig:prebiasdistance}.  The biases and observed RMS
grow with distance, as expected, approaching 30 to 40~ms at 20~kpc.

\subsection{\label{sec:after} After bias correction}

\begin{figure}
    \includegraphics[width=3.4in]{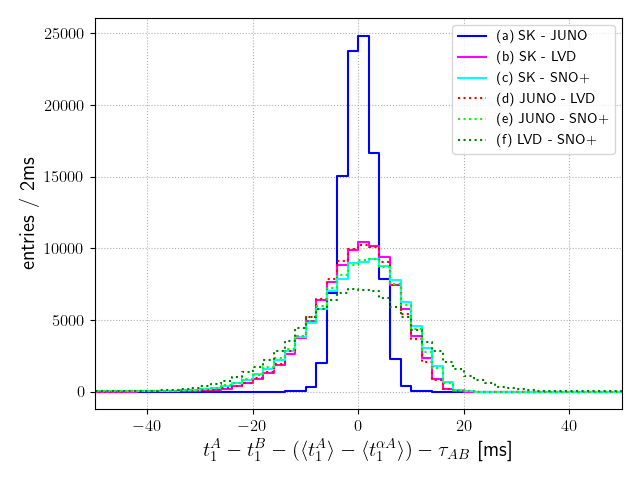}
    \caption{After bias correction:  residual bias (corrected lag less true lag, $Z_{AB}-\tau_{AB}$)
    distributions over 100,000 trials
    with the $27M_\odot$ IBD lightcurve shape at 10~kpc.}
    \label{fig:pairlags}
\end{figure}

The effect of estimating and correcting for the first-event bias is shown in the
$Z_{AB}-\tau_{AB}$ distributions of Fig.~\ref{fig:pairlags}.
As $Z_{AB}-\tau_{AB}$ is the difference between the corrected lag
and the true lag, it should ideally be zero.
Indeed, the distributions over the Monte Carlo trials are seen to be largely symmetric and centered close to zero, though a negative tail is evident in pairings between
larger and smaller detectors.
The widths of the distributions of $Z_{AB}-\tau_{AB}$ are listed in Table~\ref{tab:widths},
and are all
narrower (though just slightly in the case of Super-Kamiokande and JUNO) than the corresponding uncorrected distributions of Fig.~\ref{fig:prelags}.
\begin{table}
        \begin{tabular}{|c|c|c|c|}
        \hline
        Detector & Detector & RMS (ms) & RMS (ms) \\
        1 & 2 & uncorr. & corr. \\
        \hline
        Super-Kamiokande & JUNO & 3.1 & 3.1 \\
        Super-Kamiokande & LVD & 8.0 & 7.8 \\
        Super-Kamiokande & SNO+ & 9.0 & 8.9 \\
        JUNO & LVD & 8.0 & 7.8 \\
        JUNO & SNO+ & 9.0 & 8.9 \\
        LVD & SNO+ & 11.7 & 11.5 \\
        \hline
        \end{tabular}
        \caption{Observed RMS widths of the distributions of\\ $Z_{AB}-\tau_{AB}$,
        with the $27M_\odot$ IBD lightcurve shape at 10~kpc.}
        \label{tab:widths}
\end{table}

\begin{figure}
    \includegraphics[width=3.4in]{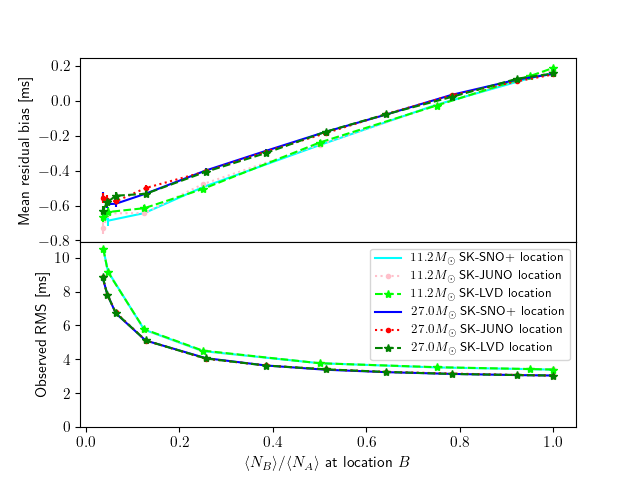}
    \caption{Top:  mean residual bias
    as a function of relative detector yield, after correcting for the first-event bias,
    for Super-Kamiokande compared with
    detector locations at JUNO, LVD, and SNO+.
    Bottom:  mean observed RMS after first-event bias correction.}
    \label{fig:resbiasrms}
\end{figure}

Fig.~\ref{fig:resbiasrms} shows the mean residual bias, {\em i.e.}, the mean of the
$Z_{AB}-\tau_{AB}$
distribution, as a function of the relative average yields.
The behavior of the mean residual bias is
largely uniform across different detector locations, and therefore across true lag values.
Even with the smallest detector, SNO+, expecting 280 events at 10 kpc from the
$27M_\odot$ model, the residual bias is less than 1~ms, compared with nearly 15~ms
before correction, an improvement better than a factor of 10.
As the yields become more symmetric, this improvement is less noticeable
(as before, the mean residual bias actually approaches $+0.2$~ms due to the
aforementioned higher energy threshold for Super-Kamiokande).
The mean observed RMS,
which reflects the measurement uncertainty for any single trial ({\it i.e.}, supernova event),
decreases from around 10~ms to a minimum variability
of around 3~ms with increasing relative yield.
For all relative yields, the residual bias is much smaller than
the observed RMS.

\begin{figure}
    \includegraphics[width=3.4in]{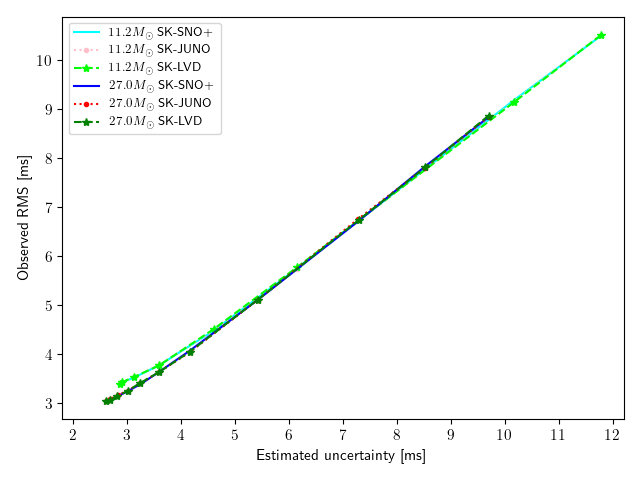}
    \caption{Mean observed RMS as a function of the mean estimated uncertainty from Eq.~\ref{eq:varz}
    for the IBD lightcurve shape.}
    \label{fig:sigmarms}
\end{figure}

Fig.~\ref{fig:sigmarms} compares
the observed RMS with the estimated uncertainty calculated using Eq.~\ref{eq:varz}.
The relationship between the observed and expected values is mostly linear, though saturating
in the $27.0M_\odot$ model at 2.6~ms for the mean of the expected uncertainties, and about 3.1~ms
for the corresponding observed RMS;
these values reflect the inherent statistical uncertainty
when comparing two detectors with the same average yields of 7800 events.
The fact that the calculated uncertainty differs from the observed variability is not surprising given the approximations used. In simplified numerical cases,
replacing $\mu(t_j)$ with $j$ by itself can be shown to account for differences of a similar size.
In order to draw reasonable confidence intervals from the data of a single supernova,
in the following we will use the estimated uncertainty and inflate it by a factor of 1.2 (the ratio between the expected and observed RMS values), even though this straightforward inflation generally overestimates the measurement uncertainty at lower yields
(see Sec.~\ref{sec:pointing}).
The corresponding values in the $11.2M_\odot$ model, with a top average yield
of 4000 events, are 3.4~ms observed and 2.9~ms
expected, also giving a ratio of 1.2.

\begin{figure}
    \includegraphics[width=3.4in]{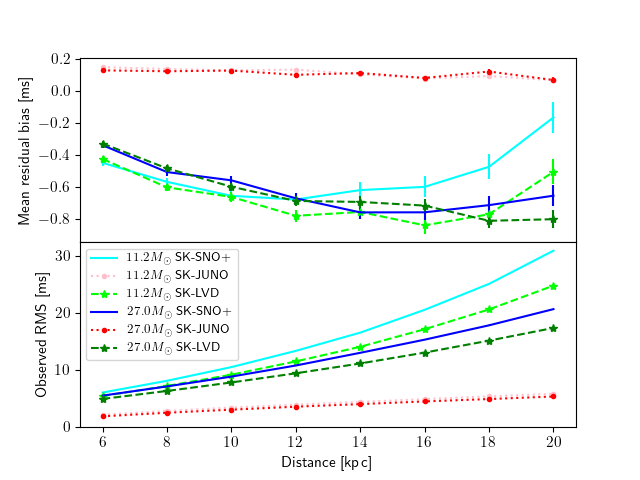}
    \caption{Mean residual bias and mean observed RMS using the IBD
    shape, but at different distances.
    }
    \label{fig:resbias20}
\end{figure}

\begin{figure}
    \includegraphics[width=3.4in]{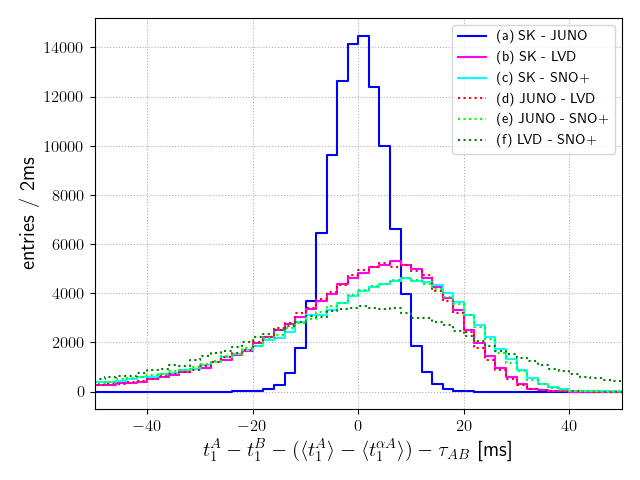}
    \caption{Residual bias distributions over 100,000 trials with the
    $27M_\odot$ IBD lightcurve shape, but at 20~kpc distance.}
    \label{fig:pairlags20}
\end{figure}

The mean residual biases and mean observed RMS are shown as a function of
supernova distance in Fig.~\ref{fig:resbias20}.
As with the uncorrected means in Fig.~\ref{fig:prebiasdistance},
the biases and observed RMS grow with distance.  The pattern changes, however, for
smaller detectors and distances above about 12~kpc.
The reason is illustrated in Fig.~\ref{fig:pairlags20}:  for pairings between a
large and a small detector, the peak of the
$Z_{AB}-\tau_{AB}$ distribution has shifted toward positive values, while a sizable
negative tail has grown, in some cases pulling the mean back to small values.
The shift is likely related to the asymmetry in the
approximation noted after Eq.~\ref{eq:nexpt1}:  while for a pair of detectors with
similar yields, such as Super-Kamiokande and JUNO,
the asymmetries largely cancel out in the difference in $Z_{AB}$,
lower yields exacerbate the asymmetry, and their effects do not
cancel when detectors of very different yields are paired. 
Indeed, the peaks of those pairings all appear on the positive side, indicating an
over-correction of the bias.

\section{Pointing}
\label{sec:pointing}

\begin{figure*}
    \includegraphics[width=2.2in]{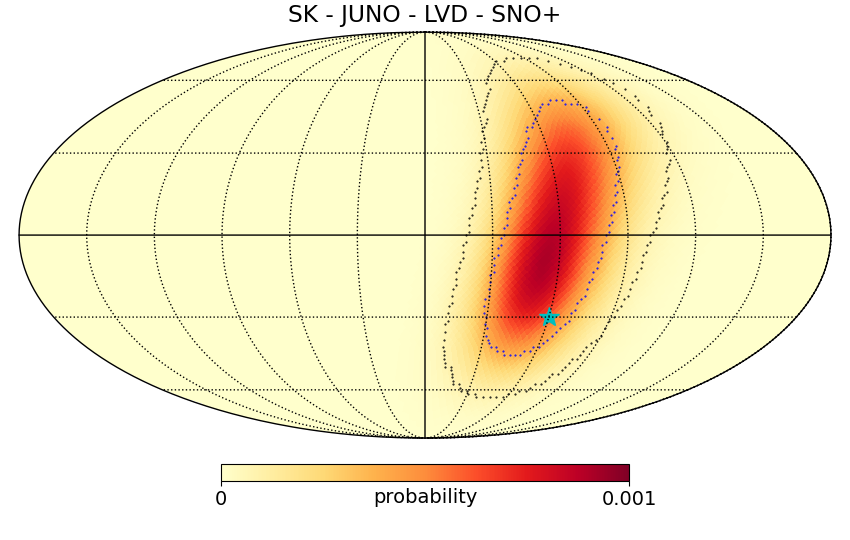}
    \includegraphics[width=2.2in]{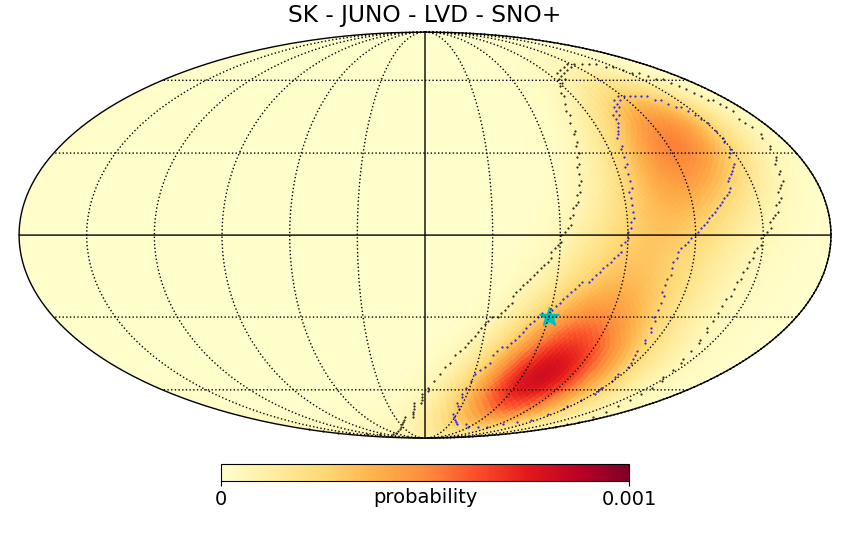}
    \includegraphics[width=2.2in]{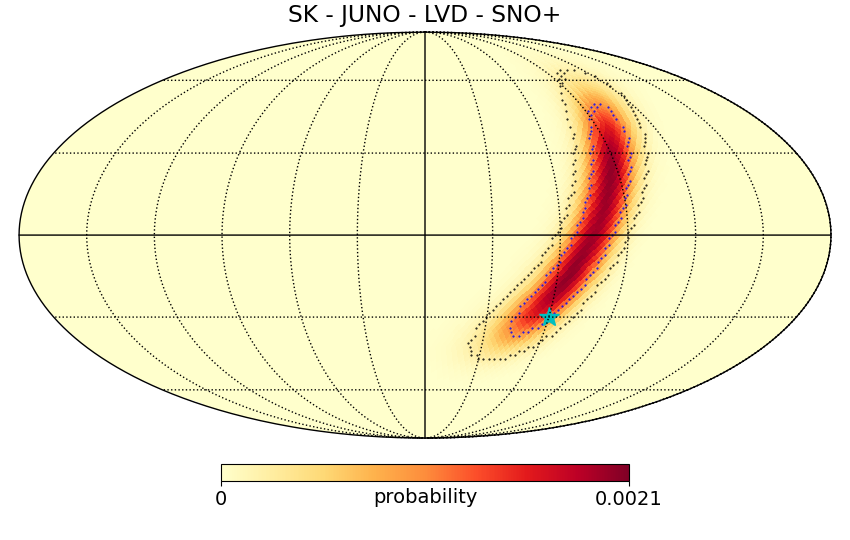}
    \caption{Sample probability skymaps for three different Monte Carlo trials combining the four detectors.  The 68\% confidence intervals are marked out by purple dots,
    with areas 3770, 5440, and 1520 square degrees, from left to right.
    The 95\% confidence intervals, marked out by black dots, cover
    8800, 11800, and 3490 square degrees.
    The true location for all the trials is indicated by the light blue star.}
    \label{fig:probmaps}
\end{figure*}

\begin{figure}
    \includegraphics[width=3.4in]{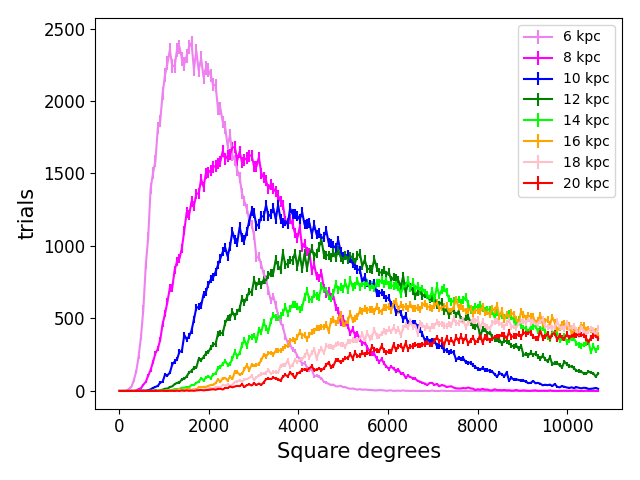}
    \caption{Coverage areas, in square degrees, of the 68\% confidence interval,
    for the $27M_\odot$ model at distances from 6 to 20~kpc.}
    \label{fig:area1sigmahist}
\end{figure}

\begin{figure}
    \includegraphics[width=3.4in]{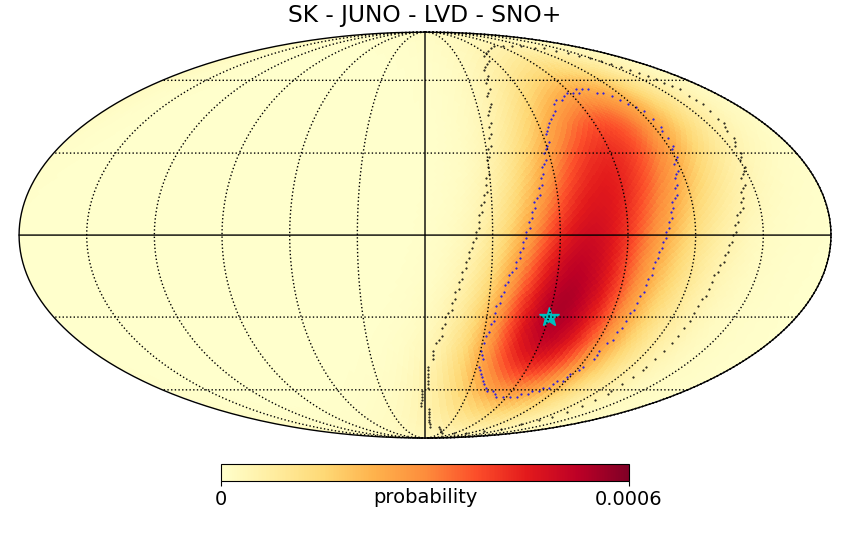}
    \caption{Probability skymap averaged over 100,000 Monte Carlo trials,
    with 68\% (purple) and 95\% (black) confidence intervals
    after first-event bias correction.
    The true location is indicated by the light blue star.}
    \label{fig:probsum}
\end{figure}

\begin{figure}
    \includegraphics[width=3.4in]{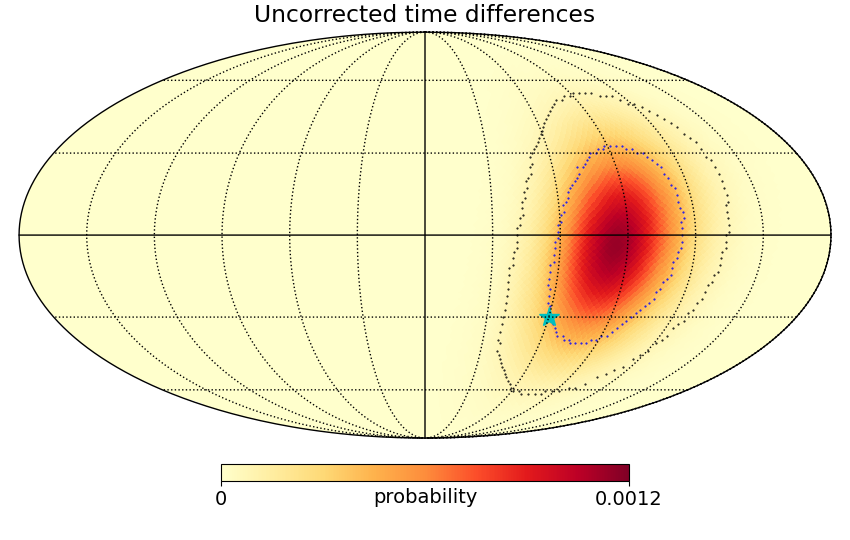}
    \caption{Probability skymap averaged over 100,000 Monte Carlo trials
    without first-event bias correction.  The true location is indicated
    by the light blue star.}
    \label{fig:probsumbad}
\end{figure}

We combine the time differences into a direction by calculating the $\chi^2$
of Eq~\ref{eq:chi2single} between
the observed time differences (with bias correction) and the predicted time differences 
for different test directions.
We use the centers of the
equal-area HEALPix~\cite{2005ApJ...622..759G}\footnote{\texttt{http://healpix.sf.net/}}
pixels, in the equatorial ICRS (or ``Celestial'') frame,
as the test directions $\uv{n}$, and assign the $\chi^2$ for that direction to each pixel.
Super-Kamiokande is designated the
reference detector $A$, and
the uncertainty estimates are inflated by the factor 1.2, as suggested in Sec.~\ref{sec:results}.
We then find the minimum $\chi^2$ value from among all the pixels and subtract
that value from all of them.  Each pixel is assigned the probability associated
with the (subtracted) $\chi^2$ value, assuming two degrees of freedom.
The resulting probability skymaps, with HEALPix $N_{side}=32$, are shown in Fig.~\ref{fig:probmaps} for three CCSN explosions simulated with the $27M_\odot$ model at a distance of 10~kpc.
As shown in Fig.~\ref{fig:area1sigmahist}, the areas covered by the 68\% confidence
interval can range from approximately 1000 to 10000 square degrees for this supernova model
and distance, with the distributions expanding and contracting with distance as expected.
An overall probability skymap for 10~kpc, averaged
over 100,000 trials, is shown in Fig.~\ref{fig:probsum}.
This skymap can be compared with Fig.~\ref{fig:probsumbad}, which has been calculated
in the same way, but without the first-event correction; in spite of the true supernova
position perching on the edge of the 68\% confidence region, it is likely that the best
observers will be looking in the wrong direction.

\begin{figure}
    \includegraphics[width=3.4in]{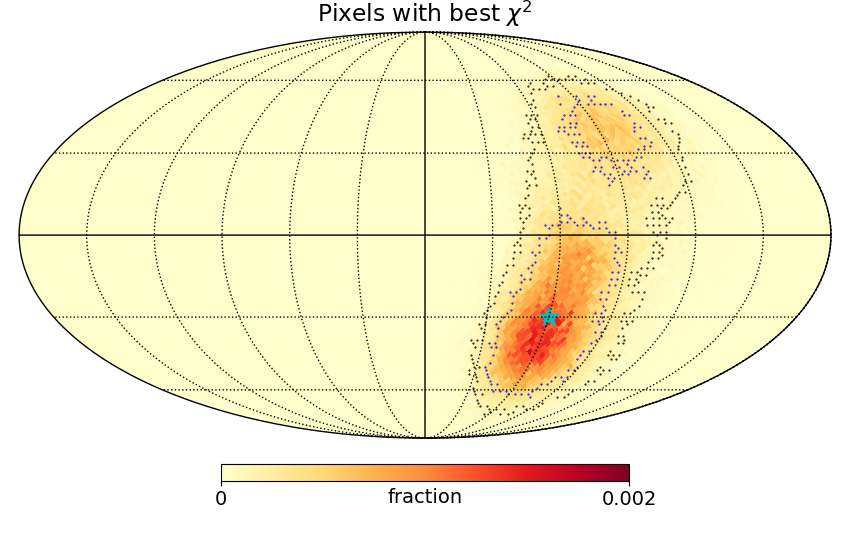}
    \caption{Skymap built with pixels of maximum probability over 100,000 trials.
    Regions including 68\% and 95\% of the best pixels are marked out by the
    purple and black dots, respectively.
    The true supernova location is indicated with the light blue star.}
    \label{fig:countmap}
\end{figure}

One way to test the coverage is to count the number of times, among the 100,000 trials, any given pixel is deemed most likely, {\em i.e.}, has the minimum $\chi^2$.
The resulting skymap, shown in Fig.~\ref{fig:countmap}, resembles the probability
maps of Figs.~\ref{fig:probmaps} and \ref{fig:probsum}, though with somewhat smaller
68\% and 95\% regions, reflecting over-coverage by the estimated uncertainties.

\begin{figure}
    \includegraphics[width=3.4in]{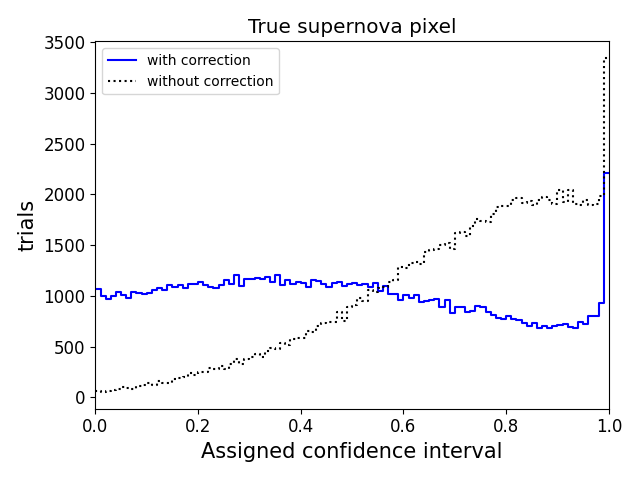}
    \caption{Confidence interval assigned to the pixel of the true supernova direction, with bias correction (blue solid) and without (black dotted).}
    \label{fig:targetcl}
\end{figure}

A more quantitative test of the coverage can be performed by calculating the confidence interval to which the true direction has been assigned. For this purpose, we plot the value of the $\chi^2$ cumulative probability distribution function for the $\chi^2$ for the pixel containing the true direction. The result is shown in Fig.~\ref{fig:targetcl}, both before and after
bias correction.
Ideally, the histogram would be flat, with (for instance)
68\% of the histogram below the value 0.68,
{\it i.e.}, with the true position lying within the 68\% region 68\% of the time. Instead, aside from a small spike near 1,
the histogram before bias correction is decidedly not flat,
with just 38\% (86\%) of the trials having the true position assigned a value
of 0.68 (0.95), while the histogram after bias correction is more flat, with
the true position lying within the purported 68\% (95\%) confidence intervals
73\% (92\%) of the time.  The spike near 1, present in both histograms,
is attributed to the effects of
non-Gaussian tails in the corresponding time difference distributions,
even after bias correction
(Figs.~\ref{fig:prelags} and \ref{fig:pairlags}).

\begin{figure}
    \includegraphics[width=3.4in]{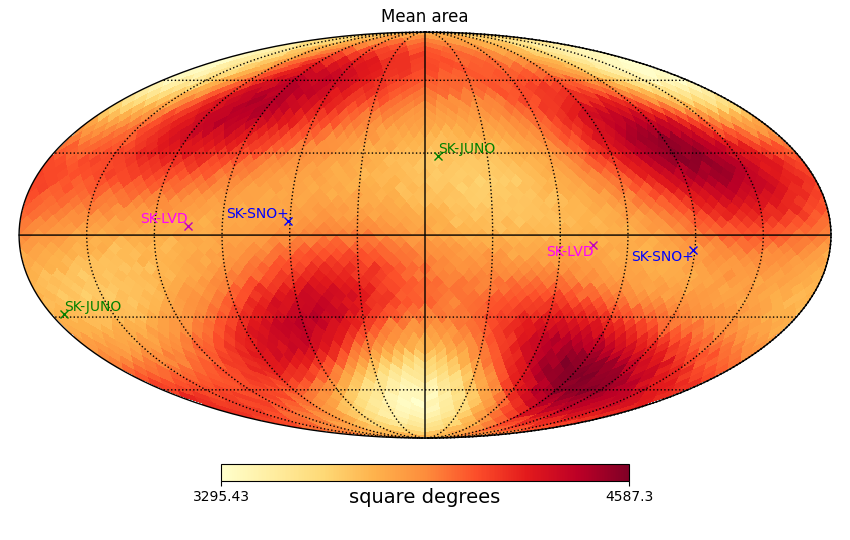}
    \caption{Average coverage areas in square degrees of the 68\% confidence interval,
    for directions defined by the centers of $N_{side}=16$ HEALPix pixels.}
    \label{fig:area1sigma}
\end{figure}

It is instructive to look at how the method works for different supernova
directions.  We place supernovae at the centers of each of the 3072 HEALPix pixels
with $N_{side}=16$, and generate 10000 Monte Carlo trials at each location.
The mean residual biases and widths are consistent across the sky.
The resulting confidence intervals, however, do vary in size with different supernova
directions due to their orientations with respect to the detectors.
The mean areas of the 68\% confidence intervals are
shown in Fig.~\ref{fig:area1sigma}, along with the directions defined by the
detector pairs at the benchmark supernova time.  The area varies from about 3300 to
4600 square degrees, with the larger areas off the detector-pair axes.
The banded structure of the skymap is evidently due to the fact that all the
detectors operate in a narrow band of latitude, from about $22^\circ$ to $46^\circ$,
all in the northern hemisphere.  The structure highlights the value of greater
geographical spread among the detectors, including, of course, detectors such as
IceCube.

\begin{figure}
    \includegraphics[width=3.4in]{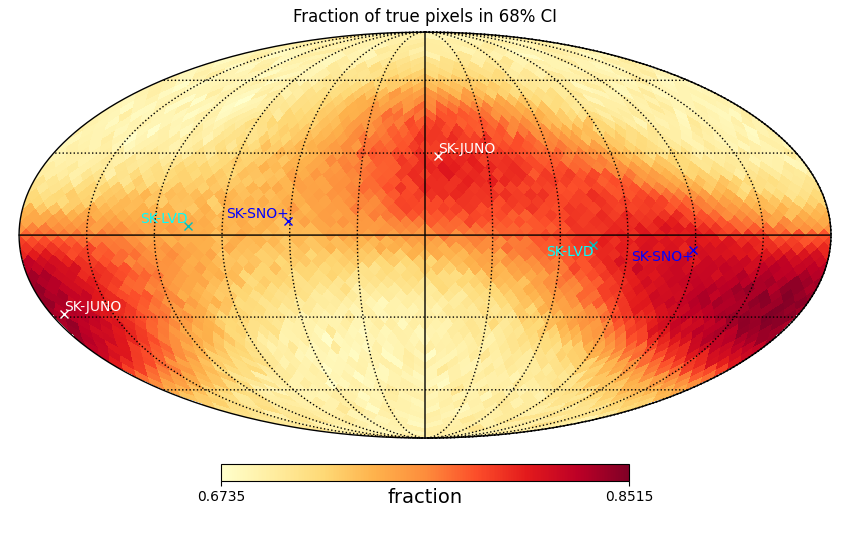}
    \caption{Fractions of the 10000 trials at each direction which are included
    in the 68\% confidence interval.}
    \label{fig:fractionsmap}
\end{figure}

\begin{figure}
    \includegraphics[width=3.4in]{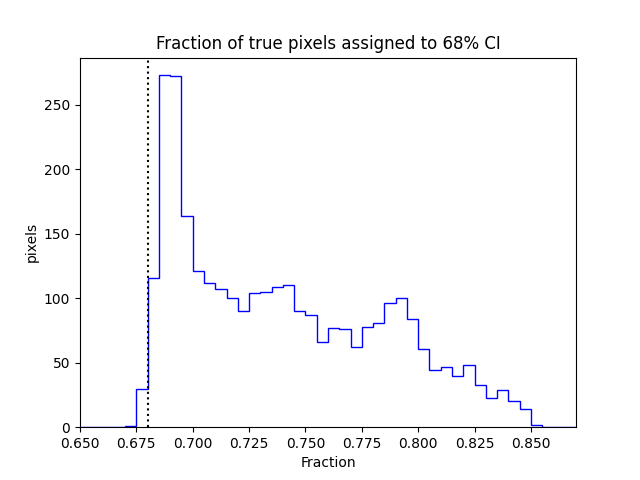}
    \caption{Histogram of fractions in Fig.~\ref{fig:fractionsmap} over
    the 3072 pixels of the $N_{side}=16$ HEALPix skymap.}
    \label{fig:fractionshist}
\end{figure}

Fig.~\ref{fig:fractionsmap} shows the 68\% confidence interval coverage in each direction,
computed in terms of the fraction of the 10000 trials in which the true direction is
included in the 68\% confidence interval.
As shown in Fig.~\ref{fig:fractionshist},
the fraction varies from 0.67 to 0.85, with the most frequent fraction around 69\%.
The behavior is consistent with the expectation that inflating the
uncertainty estimate by the factor 1.2 leads
to largely appropriate, but sometimes conservative, confidence intervals.  The
nominal supernova direction used in Sec.~\ref{sec:results} and this section
happened to point to a region of the sky with slightly conservative intervals, with
(as noted above)
the true pixel lying in the 68\% interval 73\% of the time.

\section{\label{sec:discussion} Discussion}

In this article, we have further developed the first-event method by deriving a data-driven approach for reducing the time difference bias due to two detectors' differing event yields.
The method requires only the times of the first events at most detectors,
along with a longer time series from one larger detector to act as a reference lightcurve.
The time delay between the first IBD events in different detectors, along with its estimated uncertainty, have then been used for the purpose of localizing the supernova via multilateration.
The method has been tested in Monte Carlo trials with the {\tt Bollig\_2016} models
in SNEWPY.
At a distance of 10~kpc, the residual bias is reduced from over 10~ms
to less than 1~ms when two detectors differ in size by
more than an order of magnitude, a pattern which holds
even when the distance is increased to 20~kpc.
We have also demonstrated how the uncertainty can be estimated directly from the data, and that the resulting probability skymaps
resemble those generated from Monte Carlo trials.
Moreover, the estimated uncertainties result in confidence intervals which
are reasonable, tending to the somewhat conservative, over all sky directions.
The 68\% confidence interval covers several thousand square degrees.
While this area is much larger than the approximately 30 to 150 square degrees
ultimately expected to be possible with elastic scatters, the present
method is suitable for a quick, real-time estimate in SNEWS2.0.

The main assumptions in this method are (i) that background can be neglected in the contributing detectors over the supernova burst timescale, (ii) that the shape of the time profile of the event rate is similar for all of them, and (iii) that the
early part of the event rate shape is characterized by a single, rapid rise. These assumptions are reasonable for coincidence-tagged IBD events in at least several
large underground water Cherenkov and liquid scintillator detectors. Where the background level is high, as in neutrino telescopes like IceCube and KM3NeT, the assumption of a single rapid rise may still lead to a well-defined
starting time similar to the expected first event time $\braket{t_1}$, such that after
background subtraction
a useful time difference can be integrated into first-event multilateration.  
A formula for using binned data with large
backgrounds 
has also been derived (Eq.~\ref{eq:nexpt1bin}).
It should be noted that the burst starting time used here is {\em not} the same as the bounce time, since calculating the latter requires more sophisticated model-based assumptions.

It is expected that SNEWS2.0 will employ multiple algorithms to multilaterate the supernova direction under different assumptions and levels of model dependence.
At least some methods should minimize model dependence, as the present method does. Moreover, the first-event
method can also be used as an initial direction which can be refined by slower but more precise methods that utilize features of the entire lightcurve.
For instance, algorithms using more information about the lightcurve can take advantage of distinctive features that may occur later in the supernova, such as the rapid flux decrease due to the formation of a black hole $O(1)$ second after the initial bounce~\cite{Brdar:2018zds,Wang:2021elf,Gullin:2021hfv}.
In this way, the global multi-messenger community can increase the chance to take full advantage of one of the most brilliant events in the galaxy.

\section{Acknowledgements}

FA and JT acknowledge support from the
Science and Technology Facilities Council (STFC), United Kingdom.
MCM acknowledges the support of the Belgian Fund for Scientific Research, the
FRS-FNRS (Fond pour le Recherche Scientifique).
KS, AH, and SB acknowledge NSF support from grant PHY-2209534,
JPK from grant PHY-2209449, and
DM from grants PHY-2209451 and AST-2206532.
EO is supported by the Swedish Research Council (Project No. 2020-00452).
Plots were generated with matplotlib~\cite{Hunter:2007}, and some results derived using the
healpy~\cite{Zonca2019} and HEALPix packages.
The data that support the findings of this article are openly
available via SNEWPY~\cite{Baxter2021,Baxter_2022} at~\cite{snewpy:2021}.

\bibliography{apssamp}

\appendix

\section{First event times for smaller experiments}
\label{app:alpha}

We can see that $\braket{t_1^\alpha} > \braket{t_1}$ for all $\alpha < 1$
by differentiating with respect to $\alpha$ as $\alpha$ decreases.
We start from the difference formula
\begin{equation}
    -\fderiv{\braket{t_1^\alpha}}{\alpha} = \lim_{\epsilon\rightarrow 0}
    \frac{\braket{t_1^{\alpha-\epsilon}}-\braket{t_1^\alpha}}{\epsilon}.
\label{eq:diff}
\end{equation}
The numerator is
\[
    \frac{\sum_{j=1}^N t_je^{-(\alpha-\epsilon)j}}{\sum_{j=1}^N e^{-(\alpha-\epsilon)j}}-
    \frac{\sum_{j=1}^N t_je^{-\alpha j}}{\sum_{j=1}^N e^{-\alpha j}}
\]
which to first order in $\epsilon$ (higher orders vanishing in the limit) is
\[
    \frac{(\sum_j t_jje^{-\alpha j})(\sum_k e^{-\alpha k})-
    (\sum_j t_je^{-\alpha j})(\sum_k ke^{-\alpha k})}{(\sum_j e^{-\alpha j})^2}.
\]
The denominator is always positive, so we concentrate on the numerator, which becomes
\begin{equation}
    \sum_{j=1}^N\sum_{k=1}^N t_j(j-k)e^{-\alpha(j+k)} =
    \sum_{j=1}^N\sum_{k=1}^N t_jw_{jk},
\label{eq:numerator}
\end{equation}
where $w_{jk}\equiv (j-k)e^{-\alpha(j+k)}$.  It is easy to see that $w_{jk}$ is antisymmetric
in its indices and $w_{jk}>0$ whenever $j>k$.  We split the sum into lower and upper
triangular regions in $(j,k)$
\[
    \sum_{j=1}^N\sum_{k=1}^{j-1} t_jw_{jk} +
    \sum_{j=1}^N\sum_{k=j+1}^N t_jw_{jk}.
\]
The terms in the second double sum can be reordered before swapping the dummy
indices:
\begin{eqnarray*}
    \sum_{j=1}^N\sum_{k=j+1}^N t_jw_{jk} & = & \sum_{k=1}^N\sum_{j=1}^{k-1} t_jw_{jk} \\
    & = & \sum_{j=1}^N\sum_{k=1}^{j-1} t_kw_{kj} \\
    & = & -\sum_{j=1}^N\sum_{k=1}^{j-1} t_kw_{jk}.
\end{eqnarray*}
The expression in Eq.~\ref{eq:numerator} is now
\[
    \sum_{j=1}^N\sum_{k=1}^N t_jw_{jk} =
    \sum_{j=1}^N\sum_{k=1}^{j-1} (t_j-t_k)w_{jk}.
\]
We know that since $j>k$ for all terms in the sum, $w_{jk}>0$.  Moreover,
$t_j-t_k>0$ because the event times are ordered in increasing time.
The expression in Eq.~\ref{eq:diff} thus is positive, and $\braket{t_1^\alpha}$
increases monotonically, as $\alpha$ decreases.  It is thus shown that when
using a single dataset to estimate the expectation values of the first event time
at smaller experiments, that expectation value trends later as the experiment
gets smaller, as one would expect.

\end{document}